\documentclass[aps,pre,preprint,superscriptaddress]{revtex4}
\usepackage{amsmath}
\usepackage{graphicx}
\usepackage{longtable}

\newfont{\logo}{logo10}

\newcommand{\bea}{\begin{eqnarray}}
\newcommand{\eea}{\end{eqnarray}}
\newcommand{\bes}{\begin{subequations}}
\newcommand{\ees}{\end{subequations}}

\begin{document}


\title{Simulation of universal optical logic gates under energy sharing collisions of Manakov solitons and fulfillment of practical optical logic criteria}


\author{M. Vijayajayanthi\footnote{ e-mail: vijayajayanthi.cnld@gmail.com}}
\affiliation{Department of Physics,
\\B. S. Abdur Rahman Crescent Institute of Science and Technology, Vandalur
\\ Chennai--600 048, India}
\author{T. Kanna\footnote{Corresponding author e-mail: kanna$\_$phy@bhc.edu.in}}
\affiliation{Nonlinear Waves Research Lab, PG and Research Department of Physics,
\\
Bishop Heber College,Tiruchirapalli--620 017, India}
\author{M. Lakshmanan\footnote{ e-mail: lakshman@cnld.bdu.ac.in}}
\affiliation{Department of Nonlinear Dynamics, School of Physics, Bharathidasan University, Tiruchirapalli--620 024, India\\}


\begin{abstract}
The universal optical logic gates, namely NAND and NOR gates, have been theoretically simulated by employing the energy sharing collision of bright optical solitons in the Manakov system, governing  pulse propagation in a highly birefringent fiber.  Further, we also realize the two input optical logic gates such as AND, OR, XOR, XNOR for completeness of our scheme.  Interestingly, our idea behind the simulation naturally satisfies all the criteria for practical optical logic which in turn displays the strength and versatility of our theoretical simulation of universal optical logic gates.   Hence, our approach paves the way for the experimentalists to create a new avenue in this direction if the energy sharing collisions of Manakov solitons are experimentally realized in the future.
\end{abstract}


\maketitle

\section{Introduction}
Photons can be used in the form of optical solitons to execute classical information processing and communication rather than electrons through a nonlinear medium, like highly birefringent fibre.
The rapid growth in modern day information theory and the limitations on semiconductor devices such as speed limits, energy losses and interconnect delays lead researchers to look for an alternate mode of technology.  One such highly desirable approach is to employ all optical devices to replace the conventional integrated circuits in electronics.  Light pulses are promising candidates due to their wide range of bandwidth, ultra high speed, low heat generation, etc \cite{nature}.  Optical logic gates serve as fundamental building blocks in such optical devices \cite{gokulan}.  These optical logic gates could be simulated by harnessing nonlinear effects in semiconductor optical amplifier \cite{durhuus,soto,han} and in photonic crystals.  Interestingly, here, all the logic operations are to be performed with light in the form of solitons, nonlinear coherent structures with remarkable stability. This overcomes the demerits of present day computers (digital electronics) such as processing speed, space, and heat dissipation, etc.  Realization  of practical optical logic (POL) gates by employing soliton interactions is a promising effort to replace the conventional classical computers, where electrons play major role.  In recent years, many research works were proposed and reported concerning the optical logic operations in different physical phenomena \cite{nakeeran,ieee,PRA,OE,NJP,JO,Nature}.  However, none of them has tried to verify whether they fulfill the criteria called POL. 


In the present work, we theoretically propose the simulation of universal optical logic gates by making use of the rigorous asymptotic analysis of a four soliton energy sharing collision which in turn concretely provides the mathematical support for the simulation of optical logic gates for the experimentalists.  In addition to this,  the input and output states of the solitons can be measured anywhere asymptotically rather than at a particular fibre distance. For demonstrative purpose, here we analyze our collision dynamics at the fibre distance $z=\pm15$ km.  Also, we propose the optical  logic gates for the standard Manakov model in which there is no fibre attenuation. Even in  the presence of fibre losses/attenuation, one of the authors of the present paper revealed that energy sharing property of the Manakov solitons is preserved under strong environmental perturbations\cite{josa}.  In our previous works, we theoretically explored the realization of optical single input gates \cite{one_gate} and two input gates such as OR gate and NOR gate \cite{rapid}.  Here, we wish to make a proposal for all other basic two input logic gates including the universal NAND gate. We emphasize that our theoretical approach naturally satisfies all the criteria for POL (Practical Optical Logic, see Sec. IV below for more details).
For this purpose, we consider the incoherent propagation of two orthogonally polarized high intense optical pulses in an elliptically birefringent fiber with high birefringence \cite{agarwal1,nail}. This type of pulse propagation is described by the following coupled nonlinear Schr\"odinger (CNLS) equations which can be represented by the system of coupled evolution equations \cite{menyuk1,menyuk2}
\bes
\bea
i(\Psi_{1\zeta}+\beta_{1x}\Psi_{1\tau})-\frac{\beta_{2}}{2}\Psi_{1\tau\tau}+\gamma(|\Psi_1|^2+B|\Psi_2|^2)\Psi_1=0,\\
i(\Psi_{2\zeta}+\beta_{1y}\Psi_{2\tau})-\frac{\beta_{2}}{2}\Psi_{2\tau\tau}+\gamma(|\Psi_2|^2+B|\Psi_1|^2)\Psi_2=0,
\eea
 \label{1}
\ees
where $\zeta$ and $\tau$ are respectively the propagation direction and normalized time, $\Psi_j$'s, $j=1,2,$ are complex slowly varying amplitudes, $\beta_{1x}$ and $\beta_{1y}$ are the inverse of the group velocities of the two modes, $\beta_{2}$ represents group velocity dispersion (GVD) and the effective Kerr nonlinearity coefficient $\gamma$ is defined as $\frac{8 n_2 \omega_0}{9 c A_{eff}}$, where $n_2$ is the nonlinear index coefficient, $\omega_0$ is the carrier frequency and $A_{eff}$  is the effective core area.  Here, we have neglected the rapidly varying coherent coupling terms \cite{menyuk2}. Furthermore, $\gamma$ and $\beta_2$ are chosen to be same for both the pulses as they are at the same wavelength.
The cross phase modulation (XPM) coupling parameter $B=\frac{2+2\sin^2\theta}{2+cos^2\theta}$, where $\theta$ is the ellipticity angle which can vary between 0 and $\pi/2$. We note that in the case of linearly birefringent fiber the XPM coefficient B takes the value 2/3 and in circularly birefringent fibers it becomes 2. In a specially fabricated elliptically birefringent fiber with ellipticity angle $\theta = 35^\circ$, one can achieve the B value as 1 \cite{menyuk2}. Also in telecommunication fibres, the birefringence is random. In that case, B can be made to be unity. Meanwhile, the linear and nonlinear polarization mode dispersion (PMD) terms will appear \cite{menyukPMD}  and therefore the coupled nonlinear Schr\"odinger equations can exactly be reduced to the Manakov system but with additional perturbative terms originating from the PMD.
For lossless fibres, after suitable transformations, the above equation (\ref{1})  can be expressed in the following dimensionless form using soliton units \cite{agarwal1,nail,menyuk2},
\bes
\bea
i \Psi_{1z}-\frac{sgn(\beta_2)}{2} \Psi_{1tt}+\mu(|\Psi_1|^2+B|\Psi_2|^2)\Psi_1=0,\\
i \Psi_{2z}-\frac{sgn(\beta_2)}{2} \Psi_{2tt}+\mu(|\Psi_2|^2+B|\Psi_1|^2)\Psi_2=0,
\eea
\label{cnls}
\ees
where the dimensionless length and retarded time are defined as $z=\frac{\zeta}{L_D}$, $t=\frac{T}{T_0}=(\tau-\tilde\beta_1\zeta)$ in which the dispersion length $L_D=\frac{T_0^2}{|\beta_2|}$, nonlinear length $L_{NL}=\frac{1}{\gamma P_0}$ and $\tilde\beta_1=\frac{1}{2}({\beta_{1x}+\beta_{1y}})$ with $T_0$ and $P_0$ being the initial width and peak power respectively, while $\mu=\frac{\gamma P_0 T_0^2}{|\beta_2|}>0$.
In the anomalous (normal) dispersion regime, $\beta_{2}<0~(>0)$, where the high (low) frequency pulses travel faster than the low (high) frequency pulses, the above coupled system of equations is referred as focusing (defocusing) CNLS equation and the fibre supports bright (dark and dark-bright) solitons.  These are consequences of the polarization modulation instability \cite{baronio}.  For the polarizing angle $\theta=35^\circ$ (for which $B=1$) in the anomalous dispersion regime, with a scaling transformation $z^{\prime}=\frac{z}{2}$, $q_j=\sqrt{\mu} \Psi_j, j=1,2,$ and dropping the prime, we get the standard Manakov model in normalized form as  \cite{man},
\bes
\bea
i q_{1z}+ q_{1tt}+2(|q_1|^2+|q_2|^2)q_1=0,\\
i q_{2z}+ q_{2tt}+2(|q_1|^2+|q_2|^2)q_2=0.
\eea
\label{manakov1}
\ees
The Manakov system (\ref{manakov1}) and its N - component generalization have been extensively studied (for details see \cite{man,hie,kannaprl,kannapre,dinda,kanapramana,epj} and references therein). The salient attribute of this Manakov system is the interesting  energy sharing collision of bright solitons as a consequence of change in the polarization vector during collision. In such energy sharing collision, the intensity of soliton in a given component can be suppressed (enhanced) while the other soliton experiences a converse effect. Contrary to this, the solitons in the second component display a reverse scenario thereby preserving the total intensity as well as the intensity of the individual component \cite{hie,kannaprl,kannapre,dinda,kanapramana,epj}.  Following this, multisoliton interactions in the Manakov system have been studied by using the Hirota's bilinearization method in Refs. \cite{kannapre,kanapramana,epj}.  Recently, bound state solitons and mutivalley dark solitons of various multicomponent CNLS type systems have been explored by employing the Darboux transformation method \cite{ling}. 
The Manakov bright solitons have been first experimentally observed in spatial domain with $AlGaAs$ planar waveguides \cite{expt}. On the other hand, this Manakov system serves as an appropriate framework for the experimental study of modulation instability (MI) and its connection with rogue waves and different kinds of breathers, in optical fibres. In particular, an experimental configuration to realize XPM coefficient with unity value and with very low polarization mode dispersion (PMD)\cite{menyukPMD} has been successfully demonstrated in references \cite{fatome} and \cite{wabnitz} for MI studies in the Manakov system featuring anomalous and normal group velocity dispersion, respectively.

Recently, optical dark rogue waves have also been experimentally observed in the Manakov model with defocusing nonlinearity \cite{millot}. Thus the Manakov solitons are suitable candidates for experimental realization and for further technological applications. Also, in Ref. \cite{josa},  it has been clearly demonstrated that this energy sharing property of the Manakov solitons is preserved even in the presence of fiber losses. There it has been shown that for B=2/3, with appropriate initial conditions, the bright solitons undergo efficient energy sharing collisions with a switching efficiency of 96\%. Another interesting and detailed study on system(\ref{cnls}) with B=2/3 \cite{yang} shows that for faster colliding velocity, similar transmission collision scenario occurs with larger (smaller) soliton becoming further larger (smaller). In our simulation procedure of logic gates, we fix a threshold level, above (below) which one shall have 1 (0) state. So, the observations in Ref.\cite{josa}  and in Ref. \cite{yang} suggest that our present approach of realizing 1 (0) logic and subsequent logic gate simulation will hold good.

\section{Schematic of four soliton collision process}
\begin{figure}[ht]
\includegraphics[width=0.5\linewidth]{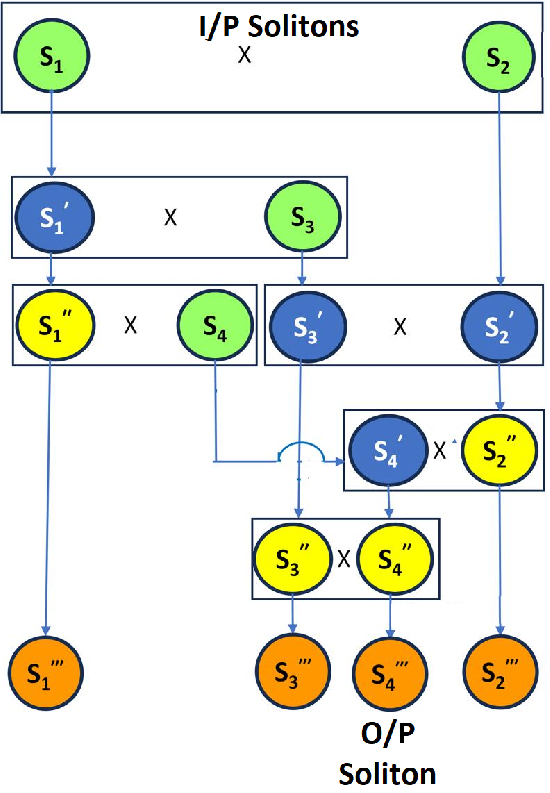}
\caption{Collision picture of four solitons $S_1$, $S_2$, $S_3$ and $S_4$.}\label{collision}
\end{figure}

In a highly birefringent fibres, it was pointed out in \cite{kannapre} that the bright Manakov solitons typically exhibit pair-wise collision under interaction.  In such a collision process, all the solitons undergo energy sharing collisions and every soliton interacts with every other solitons which are involved in the collision process.  To be specific, we employ just a four bright soliton collision process, in which each soliton undergoes three pair-wise energy sharing interactions.  One can refer from the supplemental material \cite{supp} of Ref. \cite{rapid} for the explicit four soliton solution and its detailed asymptotic analysis. We denote the input solitons as unprimed solitons $S_{j}, j=1,2,3,4$. We will refer the solitons emerging after the first, second and third pair-wise collisions as primed, double primed and triple primed solitons respectively. In fact $S_{j}^{'''}$ represent the output solitons.
The schematic pair-wise four soliton collision process considered in our present work is shown in Fig. \ref{collision}.
Here, we use the pair-wise energy sharing collisions of bright Manakov solitons as such without imposing any constraints on the colliding solitons for realizing the universal logic gate.
The intensities of the four colliding solitons at the input and at the output are calculated analytically from a systematic but rather lengthy asymptotic analysis.

\section{Notion behind the simulation of optical logic gates}
During the energy sharing collision, the Manakov solitons experience a change in their states (polarizations) due to the enhancement or suppression of intensity which is a desirable property for performing computation. Also, it is sufficient to examine these states well before (i.e., at the input) and well after (output) collisions.
 The key idea is to define the asymptotic states of the $j^{th}$ soliton as
$\rho^{j\pm}=\frac{q_1^j(z\rightarrow\pm\infty)}{q_2^j(z\rightarrow\pm\infty)}=\frac{A_1^{j\pm}}{A_2^{j\pm}},$ where $A_{1,2}^{j\pm}$ are the polarization components (1,2) of the $j^{th}$ soliton. Here subscripts denote the components, $+(-)$ designates the state after (before) collision and superscript $j$ represents the soliton number. The logic gates deal with binary logic, either $1$ or $0$. We define such a $1(0)$ state if the intensity $|\rho^{j_{\pm}}|^{2}$ of the state vector exceeds (falls below) a particular reference value.  For simulating the two input logic gate, the inputs are fed into the solitons $S_1$ and $S_2$ well before interaction, and the output is taken from the soliton  $S_4$ after interaction. The explicit forms of the states of the solitons $S_1$ and $S_2$ before interaction are
\bea
\rho^{1-}&=&\frac{\alpha_1^{(1)}}{\alpha_1^{(2)} },\\
\rho^{2-}&=&\frac{A_1^{2-}}{A_2^{2-}}=\frac{N_1^{2-}}{N_2^{2-}}=\frac{\alpha_1^{(1)} \kappa_{21}-\alpha_2^{(1)} \kappa_{11}}{\alpha_1^{(2)} \kappa_{21}-\alpha_2^{(2)} \kappa_{11}},
\eea
where
\bea
\kappa_{il}= \frac{\sum_{n=1}^2\alpha_i^{(n)}\alpha_l^{(n)*}}
{\left(k_i+k_l^*\right)},\;i,l=1,2,3,4.\nonumber
\eea
Similarly, the state of the soliton $S_4$ after collision is given by
$\rho^{4+}=\frac{\alpha_4^{(1)}}{\alpha_4^{(2)} }$.
In the above equations,  $\alpha_{l}^{(m)}$, $l=1,2,4, m=1,2,$ represent the polarization parameters of solitons $S_1$, $S_2$ and $S_4$ and they can take any arbitrary complex value.  The other quantities $\kappa_{11}, \kappa_{21}$, and $\kappa_{12}$ are defined by the soliton parameters $\alpha$'s and $k$'s.  Though the third soliton $S_3$ does not explicitly appear in the above expressions, it indirectly influences the energy sharing collisions. The ratio of intensities of two components for the solitons $S_1$ and $S_2$ before interaction as well as that for soliton $S_4$ after interaction can be obtained by taking the absolute squares of these complex states and they are given by $|\rho^{1-}|^2$, $|\rho^{2-}|^2$ and  $|\rho^{4+}|^2$, respectively.  Hence, one can measure the ratio of the intensities of the two components for the input/output solitons analytically from the asymptotic analysis. Then as mentioned before if the ratio of intensities of two components for a given  soliton $S_j$ is greater (less) than some specific threshold value, say 1, before interaction then we denote the  input state of that soliton $S_j$ as $``1 (0)"$ state.  Thus the $1$($0$) state of a particular soliton $S_j$ corresponds to $|\rho^{j-}|^2>1 (<1)$.
\section{Criteria for practical optical logic and their fulfillment }
\begin{figure}
\begin{center}
\includegraphics[width=1\linewidth]{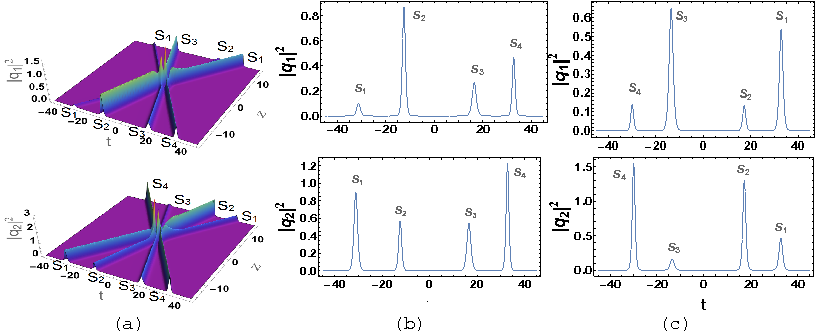}
\caption{Fanout gate: The state of input soliton ($S_1$) is $``0"$ and the state of output solitons ($S_2$ and $S_4$) are also $``0"$.  The first column (a) displays the mesh plots of the intensity profiles while the middle and last columns (b) and (c) depict the two dimensional plots of intensities at the input $(z=-15)$ and at the output $(z=15)$, respectively. }\label{fanout0}
\end{center}
\end{figure}

\begin{figure}
\begin{center}
\includegraphics[width=1\linewidth]{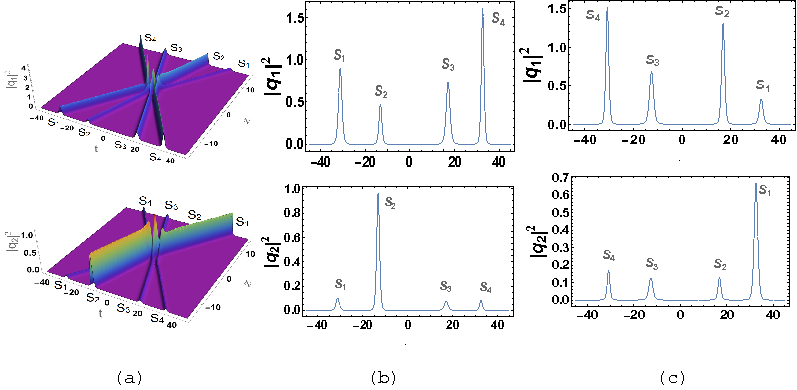}
\caption{Fanout gate: The state of input soliton ($S_1$) is $``1"$ and the state of output solitons ($S_2$ and $S_4$) are also $``1"$.  The first column (a) displays the mesh plots of the intensity profiles while the middle and last columns (b) and (c) depict the two dimensional plots of intensities at the input $(z=-15)$ and at the output $(z=15)$, respectively. }\label{fanout1}
\end{center}
\end{figure}

The article by Miller \cite{miller} lays out nicely the relevant criteria for optical logic, such as cascadability, fan-out, logic-level restoration, input/output isolation, critical biasing, and independence of loss. All these requirements are naturally met out by the proposed soliton collision based computing. As a result, the scheme does stand out in a way against a myriad of other nonlinear switching schemes.  Below, we discuss all the criteria in detail, in the framework of soliton collision.
\begin{itemize}
\item Fanout:\\
\indent In a four soliton collision process, if the input state is assigned to the soliton 1 ($S_1$) before collision, then it can be switched to the output of any of the other two solitons after collision
, say solitons $S_2$ and $S_4$, by appropriately imposing conditions on the soliton parameters.  This essentially implies the process of fanout which indicates the state of one soliton ($S_1$) before collision is used to drive as an input to at least two different solitons ($S_2$ and $S_4$) after collision.  To facilitate our understanding, we consider the following example.  The parametric values to drive the $0$ input state to solitons $S_2$ and $S_4$ are chosen as
$k_1 = 1 + i, k_2 = 1.2 + 0.5 i, k_3 = 0.9 - 0.5 i,  k_4 =
 1.3 - i, \alpha_1^{(1)} = 2, \alpha_1^{(2)} = 6, \alpha_2^{(2)} =
 1 - i, \alpha_3^{(1)} = 2, \alpha_3^{(2)} = 1;  \alpha_4^{(2)} = 2 - i$.
The parametric values of $\alpha_4^{(1)}$ and $\alpha_2^{(1)}$ are determined as $0.6-0.3 i$ and $0.002-0.4 i$, respectively, from the following conditions:
\bea
\alpha_4^{(1)}=\frac{\alpha_1^{(1)}\alpha_4^{(2)}}{\alpha_1^{(2)}},\\
\frac{N_1^{2+}}{N_2^{2+}}-\frac{\alpha_4^{(1)}}{\alpha_4^{(2)}}=0.
\eea
The expressions for $N_1^{2+}$ and $N_2^{2+}$ are already given in the asymptotic analysis(see,the supplemental material \cite{supp} of Ref. \cite{rapid}).  In order to drive the $1$ input state of the soliton $S_1$ to the solitons $S_2$ and $S_4$, we have to choose the soliton $S_1$ parameters as $\alpha_1^{(1)} = 6, \alpha_1^{(2)} = 2$ . From the above conditions, the parameters $\alpha_4^{(1)}$ and $\alpha_2^{(1)} $ are evaluated as $6-3 i, 3.4-0.8 i$  and all the other parameters are as same as for the $``0"$ state.  The fanning out of input is clearly given in the following tables I and II.  Figures \ref{fanout0} and \ref{fanout1} depict the fanning out of input states $``0"$ and $``1"$ , respectively.

\begin{table}[!ht]
\begin{minipage}{.5\linewidth}
\caption{Truth table}
\centering
\begin{tabular}{|c|c|c|}
\hline
 Input $S_1$ & Output 1 $S_{2}'''$ & Output 2 $S_{4}'''$ \\\hline
  0& 0 & 0  \\ \hline
   1& 1& 1  \\ \hline
\end{tabular}
\end{minipage}%
\begin{minipage}{.5\linewidth}
\centering
\caption{Intensity table }
\begin{tabular}{|c|c|c|}
\hline
  Input $S_1$ & Output 1 $S_{2}'''$ & Output 2 $S_{4}'''$ \\\hline
  0.1& 0.1 & 0.1  \\ \hline
   9& 9& 9  \\ \hline
\end{tabular}
\end{minipage}
\end{table}

\item Logic-level restoration:\\
\indent Here the optical pulses are propagating in the form of soliton which is by nature a localized coherent structure that can travel over long distances without alteration in shape and robust against strong environmental perturbations.  This special property of solitons can restore the logic signal throughout its propagation in an optical fibre.

\item Input/output Isolation:\\
\indent  In the realization of optical logic gates, the inputs are fed into the first two pulses before collision, say soliton  $S_1$ and soliton  $S_2$, respectively.  Then the output is measured in the last pulse after collision(totally four solitons involved in the composite collision), say soliton  $S_4$. This final soliton is well separated from the input soliton. Hence, the input and output solitons are treated separately which will prevent the input pulses to be reflected back into the output pulse.

\item Logic level independent of loss:\\
\indent To achieve this, we need a differential signaling that requires the ratios of powers or difference of powers at the input.  Interestingly, in our present work, indeed we define the state of a particular soliton as the ratio of the intensities of the soliton propagating in the two components. For example,
\bea
|\rho^{1}|^2=\frac{|A_1^{1}|^2}{|A_2^{1}|^2},
\eea
where $|\rho^{1}|^2$ represents the state of the soliton $S_1$ , $|A_1^{1}|^2$ represents the intensity of the soliton $S_1$ propagating in $q_1$ mode and $|A_2^{1}|^2$ denotes the intensity of the soliton $S_1$  propagating in $q_2$ mode.  Thus, we have taken the ratio of intensities of the solitons propagating in two different modes ($q_1$ and $q_2$) to represent the state of a particular soliton, say soliton $S_1$.  This automatically satisfies the differential signaling criteria in \cite{miller}.

\item Absence of critical biasing:\\
\indent In order to achieve  ``1"(``0") states, we do not require a precise value of ratio of intensities of two components for the solitons. Rather, one can fix an arbitrary threshold value above (below) which all the values of itensities can be treated as  ``1"(``0") states.  Specifically, we are considering ``1"(``0") states which are having $|\rho|^2$ values above (below) a threshold. And hence, there is no need for high precision for the states.  By this way, we introduce the absence of critical biasing in the soliton collision based optical computing.
\item Cascadability:\\
\indent We employ the four soliton collision process for the simulation of NOR gate in the present manuscript.  However, in our earlier work \cite{one_gate}, we have demonstrated the simulation of one input gates such as COPY gate, NOT gate and ONE gate by employing a three soliton collision process. Here the input is fed into the second soliton $S_2$ before collision and the output is taken up from the third soliton $S_3$ after collision.

In the present paper, it will be shown that for the simulation of NOR gate, the inputs are fed into the solitons $S_1$ and $S_2$ before collision and the output is taken up from soliton $S_4$ after collision.  Instead, in order to use the criteria of cascadability, one can use the output of the one input gate say COPY gate(which can be realized from the three soliton collision) $S_3''$  as one of the inputs to the universal NOR gate whereas the first soliton $S_1$ can be treated as another input, while the output is taken up from soliton $S_4$ as usual.  In a nutshell, we may claim that one can simulate the two input gate (universal gate) by cascading the output of the one input gate.
\end{itemize}
Thus, the fulfillment of POL by our present proposed work and the experiments on Manakov solitons and their energy sharing collisions \cite{akhmediev,wabnitz} demonstrate the definite possibility to simulate optical logic gates satisfying POL.   This is an important advancement in soliton collision based computing.
\section{Simulation of two input optical logic gates using four soliton collisions}
In this section, we demonstrate the simulation of all the two input optical logic gates including universal gates such as NAND, NOR, AND, OR, XOR, XNOR gates from out of the four soliton collision process.
\subsection{NAND gate}
\begin{figure}
\begin{center}
\includegraphics[width=1\linewidth]{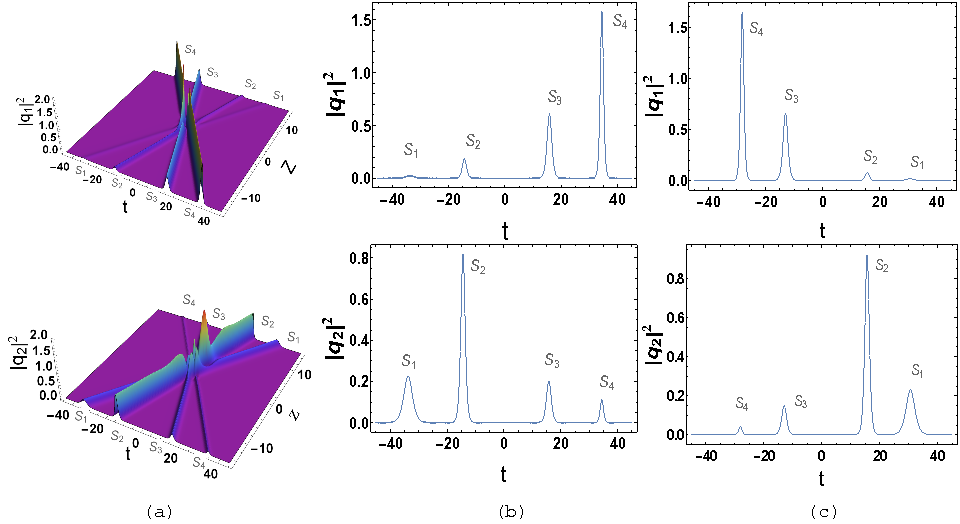}
\caption{NAND gate: The state of input solitons ($S_1$ and $S_2$) are $``0"$ and  $``0"$ and the state of output soliton ($S_4$) is $``1"$.  The first column (a) displays the mesh plots of the intensity profiles while the middle and last columns (b) and (c) depict the two dimensional plots of intensities at the input $(z=-15)$ and at the output $(z=15)$, respectively. }\label{nand_00}
\end{center}
\end{figure}

To achieve the required output corresponding to the NAND gate, we deduce the following condition on the soliton parameters from asymptotic analysis:
\bea
\hspace{-0.3cm}\alpha_4^{(2)}=\left(\frac{\alpha_1^{(1)}}{\alpha_1^{(2)}}\times\frac{\alpha_2^{(1)}\left((k_1-k_2)|\alpha_1^{(1)}|^2-(k_2+k_1
^*)|
\alpha_1^{(2)}|^2\right)+(k_1+k_1^*)\alpha_1^{(1)}\alpha_1^{(2)*}\alpha_2^{(2)}}{(k_1+k_1^*)\alpha_1^{(1)*}\alpha_2^{(1)}\alpha_1^{(2)}-\alpha_2^{(2)}\left((k_2-k_1)
|\alpha_1^{(2)}|^2+(k_2+k_1^*)|\alpha_1^{(1)}|^2\right)}\right) \alpha_4^{(1)}.
\label{con1}
\eea
The above relation (\ref{con1}) is obtained by imposing the condition, $\rho^{4+}=\left(\rho^{1-} \rho^{2-}\right)^{-1}$ on the state vectors of the input solitons ($S_1$, $S_2$) and the output soliton ($S_4$) so that the Boolean algebra of the NAND gate is satisfied. Assigning $(0,0)$ input states to ($S_1$, $S_2$) by choosing $\alpha_1^{(1)}=2,  \alpha_1^{(2)}=6, \alpha_2^{(1)}=2, $ and $\alpha_2^{(2)}=5,$ we achieve the $``1"$ output state from soliton $S_4$ for the parameter choices $k_1=0.5+i, k_2=1+0.5 i, k_3=0.9-0.5 i, k_4=1.3-i, \alpha_3^{(1)}=3, \alpha_3^{(2)}=1, \alpha_4^{(1)}=0.1-0.2 i$ along with the condition (\ref{con1}), which is depicted in Fig \ref{nand_00}.
Assigning $(0,1)$ input states to ($S_1$, $S_2$) by choosing $\alpha_1^{(1)}=2,  \alpha_1^{(2)}=6, \alpha_2^{(1)}=50, \alpha_2^{(2)}=45,$ we achieve the $``1"$ output state from soliton $S_4$.  All the other soliton parameters are chosen as same as above for all the combinations of input states.
The $(1,0)$ input states are fed into the solitons ($S_1$, $S_2$) by choosing $\alpha_1^{(1)}=11,  \alpha_1^{(2)}=2, \alpha_2^{(1)}=2, \alpha_2^{(2)}=5,$ so that we achieve the $``1"$ output state from soliton $S_4$.
Assigning $(1,1)$ input states to ($S_1$, $S_2$) by choosing $\alpha_1^{(1)}=9,  \alpha_1^{(2)}=2, \alpha_2^{(1)}=7, \alpha_2^{(2)}=1.5,$ we achieve the $``0"$ output state from soliton $S_4$.
\begin{table}
\begin{minipage}{.5\linewidth}
\caption{Truth table of NAND gate}
\begin{tabular}
{|c|c|c|c|}
\hline
 Input 1  & Input 2  & Output \\
 ($S_1$) & ($S_2$) & ($S_{4}'''$)  \\\hline
0& 0 & 1 \\ \hline
   0& 1& 1 \\ \hline
1& 0 & 1\\ \hline
1& 1 & 0  \\ \hline
\end{tabular}
\end{minipage}%
\begin{minipage}{.5\linewidth}
\centering
\caption{Intensity table of NAND gate}
\begin{tabular}{|c|c|c|c|}
\hline
Input 1 \;\; & Input 2 \;\;& Output \;\;\\
$|\rho^{1-}|^2$   &$|\rho^{2-}|^2$   &$|\rho^{4+}|^2$ \\ \hline
 0.1& 0.2 &33 \\ \hline
   0.1& 7& 1.6 \\ \hline
24& 0.02 & 1.8  \\ \hline
23& 32 & 0.005 \\ \hline
\end{tabular}
\end{minipage}
\end{table}
The truth table and the corresponding intensity tables (calculated values of the ratios of intensities of solitons) are given in Tables III and IV.  The other universal gate, namely NOR gate has already been demonstrated in our recent work \cite{rapid}.
\subsection{AND gate}
\begin{figure}
\begin{center}
\includegraphics[width=1\linewidth]{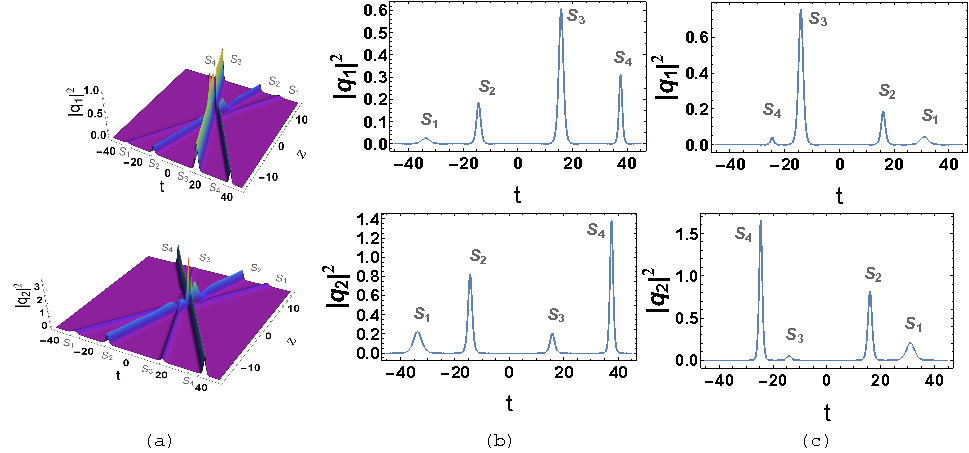}
\caption{AND gate: The state of input solitons ($S_1$ and $S_2$) are $``0"$ and  $``0"$ and the state of output soliton ($S_4$) is $``0"$.  The first column (a) displays the mesh plots of the intensity profiles while the middle and last columns (b) and (c) depict the two dimensional plots of intensities at the input $(z=-15)$ and at the output $(z=15)$, respectively.}\label{and_00}
\end{center}
\end{figure}

As discussed in the NAND gate, here we deduce the following condition on the soliton parameters from asymptotic analysis in order to achieve the required output corresponding to the AND gate:
\bea
\hspace{-0.3cm}\alpha_4^{(1)}=\left(\frac{\alpha_1^{(1)}}{\alpha_1^{(2)}}\times\frac{\alpha_2^{(1)}\left((k_1-k_2)|\alpha_1^{(1)}|^2-(k_2+k_1^*)|
\alpha_1^{(2)}|^2\right)+(k_1+k_1^*)\alpha_1^{(1)}\alpha_1^{(2)*}\alpha_2^{(2)}}{(k_1+k_1^*)\alpha_1^{(1)*}\alpha_2^{(1)}\alpha_1^{(2)}-\alpha_2^{(2)}\left((k_2-k_1)
|\alpha_1^{(2)}|^2+(k_2+k_1^*)|\alpha_1^{(1)}|^2\right)}\right) \alpha_4^{(2)}.
\label{con2}
\eea
The above relation (\ref{con2}) is obtained by imposing the condition, $\rho^{4+}=\rho^{1-} \rho^{2-}$ on the state vectors of the input solitons ($S_1$, $S_2$) and the output soliton ($S_4$) so that the Boolean algebra of the AND gate is satisfied.  Now, assigning $(0,0)$ input states to ($S_1$, $S_2$) by choosing $\alpha_1^{(1)}=2,  \alpha_1^{(2)}=6, \alpha_2^{(1)}=2, \alpha_2^{(2)}=5,$ we achieve the $``0"$ output state from soliton $S_4$ for the parameter choices $k_1=0.5+i, k_2=1+0.5 i, k_3=0.9-0.5 i, k_4=1.3-i, \alpha_3^{(1)}=3, \alpha_3^{(2)}=1, \alpha_4^{(2)}=0.001-0.002 i$ along with the condition (\ref{con2}), which is depicted in Fig \ref{and_00}.  For the $(0,1)$ input states of the solitons ($S_1$, $S_2$),  we obtain the $``0"$ output state from the soliton $S_4$ by choosing $\alpha_1^{(1)}=2,  \alpha_1^{(2)}=6, \alpha_2^{(1)}=50, \alpha_2^{(2)}=45$. All the other soliton parameters are chosen as same as above for all other combinations of input states.
Assigning $(1,0)$ input states to ($S_1$, $S_2$) by choosing $\alpha_1^{(1)}=6,  \alpha_1^{(2)}=2, \alpha_2^{(1)}=2, \alpha_2^{(2)}=5,$ we achieve the $``0"$ output state from soliton $S_4$.
Finally, $(1,1)$ input states are given to the solitons ($S_1$, $S_2$) by choosing $\alpha_1^{(1)}=9,  \alpha_1^{(2)}=2, \alpha_2^{(1)}=7, \alpha_2^{(2)}=1.5,$ and we obtain the $``1"$ output state from soliton $S_4$.
The truth table and the corresponding intensity tables are given in tables V and VI. The detailed demonstration of the OR logic gate is discussed in the Supplemental Material of our recent work \cite{supp}.
\begin{table}
\begin{minipage}{.5\linewidth}
\caption{Truth table of AND gate}
\begin{tabular}
{|c|c|c|c|}
\hline
 Input 1  & Input 2  & Output \\
 ($S_1$) & ($S_2$) & ($S_{4}'''$)  \\\hline
0& 0 & 0 \\ \hline
   0& 1& 0 \\ \hline
1& 0 & 0\\ \hline
1& 1 & 1  \\ \hline
\end{tabular}
\end{minipage}%
\begin{minipage}{.5\linewidth}
\centering
\caption{Intensity table of AND gate}
\begin{tabular}{|c|c|c|c|}
\hline
Input 1 \;\; & Input 2 \;\;& Output \;\;\\
$|\rho^{1-}|^2$   &$|\rho^{2-}|^2$   &$|\rho^{4+}|^2$ \\ \hline
 0.1& 0.2 &0.03 \\ \hline
   0.1& 6.9& 0.6 \\ \hline
7.6& 0.02 & 0.2  \\ \hline
23& 21 & 165 \\ \hline
\end{tabular}
\end{minipage}
\end{table}

\subsection{XOR gate}
\begin{figure}
\begin{center}
\includegraphics[width=1\linewidth]{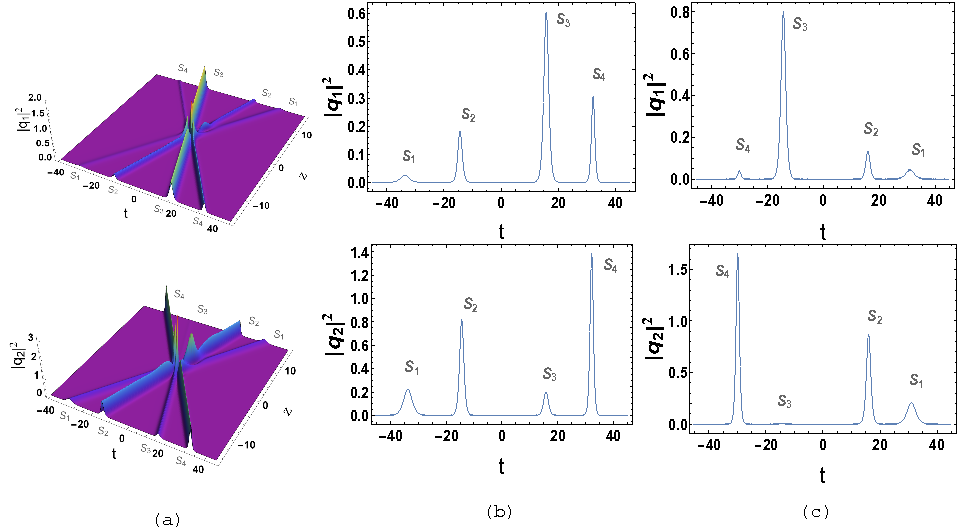}
\caption{XOR gate: The state of input solitons ($S_1$ and $S_2$) are $``0"$ and  $``0"$ and the state of output soliton ($S_4$) is $``0"$.  The first column (a) displays the mesh plots of the intensity profiles while the middle and last columns (b) and (c) depict the two dimensional plots of intensities at the input $(z=-15)$ and at the output $(z=15)$, respectively. }\label{ex_or_00}
\end{center}
\end{figure}

To simulate the XOR gate,  two inputs are fed into the solitons $S_1$ and $S_2$ and the output is taken up from soliton $S_4$, as usual. In order to get the desired output satisfying the truth table of the XOR gate, we make use of the condition $\rho^{4+}=\rho^{1-}-\rho^{2-}$ and choose
\bea
\alpha_4^{(1)}= \left(\frac{\alpha_1^{(1)}}{\alpha_1^{(2)}}-\frac{\alpha_2^{(1)}\left((k_1-k_2)|\alpha_1^{(1)}|^2-(k_2+k_1^*)|
\alpha_1^{(2)}|^2\right)+(k_1+k_1^*)\alpha_1^{(1)}\alpha_1^{(2)*}\alpha_2^{(2)}}{(k_1+k_1^*)\alpha_1^{(1)*}\alpha_2^{(1)}\alpha_1^{(2)}-\alpha_2^{(2)}\left((k_2-k_1)
|\alpha_1^{(2)}|^2+(k_2+k_1^*)|\alpha_1^{(1)}|^2\right)}\right) \alpha_4^{(2)}.
\label{con3}
\eea
Assigning the $(0,0)$ input states to ($S_1$, $S_2$) by choosing $\alpha_1^{(1)}=2,  \alpha_1^{(2)}=6, \alpha_2^{(1)}=2, \alpha_2^{(2)}=5,$ we achieve the $``0"$ output state from soliton $S_4$ for the parameter choices $k_1=0.5+i, k_2=1+0.5 i, k_3=0.9-0.5 i, k_4=1.3-i, \alpha_3^{(1)}=3, \alpha_3^{(2)}=1, \alpha_4^{(2)}=2-i$ along with the condition (\ref{con3}), which is shown in Fig \ref{ex_or_00}.
Like wise, setting the $(0,1)$ input states to ($S_1$, $S_2$) by choosing $\alpha_1^{(1)}=2,  \alpha_1^{(2)}=6, \alpha_2^{(1)}=5, \alpha_2^{(2)}=2,$ we achieve the $``1"$ output state from soliton $S_4$ for the above parameter choices.  All the other soliton parameters are as same as mentioned above for all the other combinations of input.
 Assiging $(1,0)$ input states to ($S_1$, $S_2$) by choosing $\alpha_1^{(1)}=6,  \alpha_1^{(2)}=2, \alpha_2^{(1)}=2, \alpha_2^{(2)}=5,$ we achieve the $``1"$ output state from soliton $S_4$ for the above parameter choices.
Finally, the $(1,1)$ input states are fed into the solitons ($S_1$, $S_2$) by choosing $\alpha_1^{(1)}=6,  \alpha_1^{(2)}=2, \alpha_2^{(1)}=5, \alpha_2^{(2)}=2,$ and we achieve the $``0"$ output state from soliton $S_4$ for the parameter choices. Tables VII and VIII provide the truth table and the corresponding intensity table of the XOR gate, respectively.
\begin{table}
\begin{minipage}{.5\linewidth}
\caption{Truth table of XOR gate}
\begin{tabular}
{|c|c|c|c|}
\hline
 Input 1  & Input 2  & Output \\
 ($S_1$) & ($S_2$) & ($S_{4}'''$)  \\\hline
0& 0 & 0 \\ \hline
   0& 1& 1 \\ \hline
1& 0 & 1\\ \hline
1& 1 & 0  \\ \hline
\end{tabular}
\end{minipage}%
\begin{minipage}{.5\linewidth}
\centering
\caption{Intensity table of XOR gate}
\begin{tabular}{|c|c|c|c|}
\hline
Input 1 \;\; & Input 2 \;\;& Output \;\;\\
$|\rho^{1-}|^2$   &$|\rho^{2-}|^2$   &$|\rho^{4+}|^2$ \\ \hline
 0.08& 0.2 &0.02 \\ \hline
   0.1& 49& 54 \\ \hline
8& 0.03 & 10  \\ \hline
7.6& 4.5 & 0.9
\\ \hline
\end{tabular}
\end{minipage}
\end{table}
\subsection{XNOR gate}
\begin{figure}
\begin{center}
\includegraphics[width=1\linewidth]{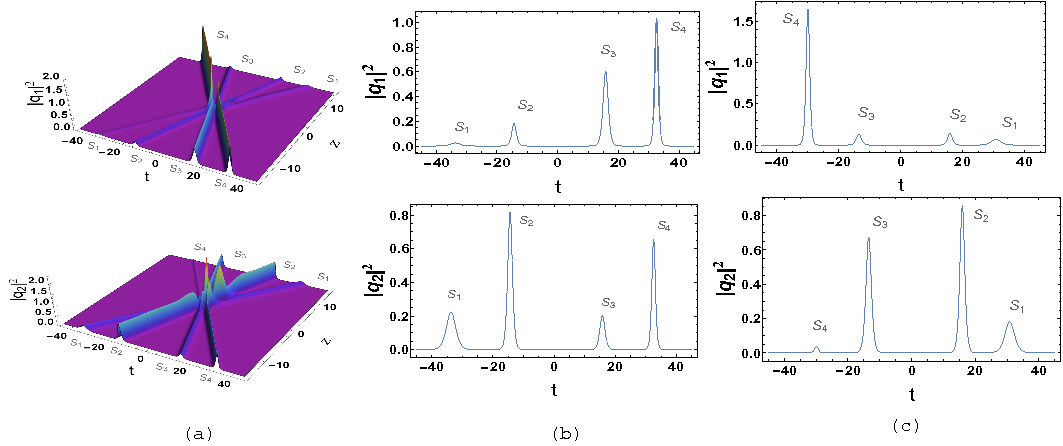}
\caption{XNOR gate: The state of input solitons ($S_1$ and $S_2$) are $``0"$ and  $``0"$ and the state of output soliton ($S_4$) is $``1"$.  The first column (a) displays the mesh plots of the intensity profiles while the middle and last columns (b) and (c) depict the two dimensional plots of intensities at the input $(z=-15)$ and at the output $(z=15)$, respectively. }\label{ex_nor_00}
\end{center}
\end{figure}

 In order to get the desired output satisfying the truth table of the XNOR gate, we make use of the condition $\rho^{4+}=\left(\rho^{1-}-\rho^{2-}\right)^{-1}$ and choose
\bea
\alpha_4^{(2)}= \left(\frac{\alpha_1^{(1)}}{\alpha_1^{(2)}}-\frac{\alpha_2^{(1)}\left((k_1-k_2)|\alpha_1^{(1)}|^2-(k_2+k_1^*)|
\alpha_1^{(2)}|^2\right)+(k_1+k_1^*)\alpha_1^{(1)}\alpha_1^{(2)*}\alpha_2^{(2)}}{(k_1+k_1^*)\alpha_1^{(1)*}\alpha_2^{(1)}\alpha_1^{(2)}-\alpha_2^{(2)}\left((k_2-k_1)
|\alpha_1^{(2)}|^2+(k_2+k_1^*)|\alpha_1^{(1)}|^2\right)}\right) \alpha_4^{(1)}.
\label{con4}
\eea
Setting the $(0,0)$ input states to ($S_1$, $S_2$) by choosing $\alpha_1^{(1)}=2,  \alpha_1^{(2)}=6, \alpha_2^{(1)}=2, \alpha_2^{(2)}=5,$ we achieve the $``1"$ output state from the soliton $S_4$ for the parameter choices $k_1=0.5+i, k_2=1+0.5 i, k_3=0.9-0.5 i, k_4=1.3-i, \alpha_3^{(1)}=3, \alpha_3^{(2)}=1, \alpha_4^{(1)}=2-i $ along with the condition (\ref{con4}), which is depicted in Fig \ref{ex_nor_00}.
Assigning the $(0,1)$ input states to ($S_1$, $S_2$) by choosing $\alpha_1^{(1)}=2,  \alpha_1^{(2)}=6, \alpha_2^{(1)}=5, \alpha_2^{(2)}=2,$ we obtain the $``0"$ output state from the soliton $S_4$ for the above parameter choices.  All the other soliton parameters are as same as mentioned above for all the other combinations of input. The
 $(1,0)$ input states are fed into the solitons ($S_1$, $S_2$) by choosing $\alpha_1^{(1)}=6,  \alpha_1^{(2)}=2, \alpha_2^{(1)}=2, \alpha_2^{(2)}=5,$ and we achieve the $``0"$ output state from the soliton $S_4$ for the parameter choices.
Finally the $(1,1)$ input states are given to the solitons ($S_1$, $S_2$) by choosing $\alpha_1^{(1)}=6,  \alpha_1^{(2)}=2, \alpha_2^{(1)}=5, \alpha_2^{(2)}=2,$ and we obtain the $``1"$ output state from the soliton $S_4$ for the above parameter choices. The truth table and the corresponding intensity tables of the XNOR gate are given in Tables IX and X, respectively.
\begin{table}
\begin{minipage}{.5\linewidth}
\caption{Truth table of XNOR gate}
\begin{tabular}
{|c|c|c|c|}
\hline
 Input 1  & Input 2  & Output \\
 ($S_1$) & ($S_2$) & ($S_{4}'''$)  \\\hline
0& 0 & 1 \\ \hline
   0& 1& 0 \\ \hline
1& 0 & 0\\ \hline
1& 1 & 1  \\ \hline
\end{tabular}
\end{minipage}%
\begin{minipage}{.5\linewidth}
\centering
\caption{Intensity table of XNOR gate}
\begin{tabular}{|c|c|c|c|}
\hline
Input 1 \;\; & Input 2 \;\;& Output \;\;\\
$|\rho^{1-}|^2$   &$|\rho^{2-}|^2$   &$|\rho^{4+}|^2$ \\ \hline
 0.13& 0.2 &41 \\ \hline
   0.14& 32& 0.02 \\ \hline
7.6& 0.03 & 0.1  \\ \hline
7& 5 & 1.1
\\ \hline
\end{tabular}
\end{minipage}
\end{table}
\section{Conclusion}
We have theoretically demonstrated the simulation of optical universal logic gates, namely, the NAND and NOR gates  using the energy sharing collision of four bright solitons in a randomly varying highly birefringent fiber described by the Manakov system. For completeness of our study, we have also simulated AND, OR, XOR, and XNOR optical logic gates.  The greatest advantage of our present work is that the notion behind the simulation of two input logic gates satisfies all the criteria for practical optical logic. One can extend the work to realize flip flops, half adder and full adder, etc.   The same theoretical realization of universal gates may be accomplished by using bound solitons in future.  Work is in progress in this direction and it paves the way to simulate multistate logic and memory storage devices. We hope that the present research work will shed more light to the experimentalists who are interested to realize energy sharing collision of solitons.
\section*{Acknowledgement}
M. V. thanks the Management, B. S. Abdur Rahman Crescent Institute of Science and Technology for the extended support.  T. K. acknowledges the support received from SERB - Department of Science and Technology, Government of India, through a Core Research Grant No. CRG/2021/004119.  M. L. wishes to acknowledge the financial support under the DST - SERB National Science Chair position awarded to him.

\end{document}